# Elucidating the Degradation Mechanism of $Gd_2Zr_2O_7$ Waste Form under Multi-Energy He Ion Irradiation

Junjing Duan [a, b], Zhangyi Huang [a, c, *], Xunxiang Hu [a], Haomin Wang [c], Yao Yang [a, b], Esra Y. Mertsoy [d, e, f], Di Wu [e, f, g, h *], Jianqi Qi [a, b, i *], Tiecheng Lu [a, b, i]

[a] College of Physics, Sichuan University, Chengdu 610064, China

[b] Key Laboratory of Radiation Physics and Technology of Ministry of Education, Sichuan University, Chengdu 610064, China

[c] Institute for Advanced Study, Chengdu University, Chengdu 610106, China

[d] Alexandra Navrotsky Institute for Experimental Thermodynamics, Washington State University, Pullman, Washington, 99164, United States

[e] The Gene and Linda Voiland School of Chemical Engineering and Bioengineering, Washington State University, Pullman, Washington, 99164, United States

[f] Department of Chemical Engineering, Çankırı Karatekin University, Çankırı, 18100, Turkey

[g] Department of Chemistry, Washington State University, Pullman, Washington, 99164, United States

[h] Materials Science and Engineering, Washington State University, Pullman, Washington, 99164, United States

[i] Key Laboratory of High Energy Density Physics of Ministry of Education, Sichuan University, Chengdu 610064, China

Corresponding Authors:

Zhangyi Huang: huangzhangyi@cdu.edu.cn

Di Wu: d.wu@wsu.edu

Jianqi Qi: qijianqi@scu.edu.cn

**Abstract**

We studied the microstructural and helium bubbling evolutions of $Gd_2Zr_2O_7$ waste form with immobilized TRPO (50 wt%) under multi-energy He ion irradiation. *Three* structurally heterogeneous regions for the $Gd_2Zr_2O_7$ waste form were found as a function of the depth from the He-irradiated surface. Specifically, at a depth less than 40 nm below the He-irradiated surface (*Region I*) the $Gd_2Zr_2O_7$ waste form is completely amorphous with large spherical He bubbles (5–25 nm). In the intermediate region, *Region II*, (40–800 nm) partially amorphized $Gd_2Zr_2O_7$ waste form accompanied with ribbon-like He bubbles that may lead to the formation of microcracks is observed. The crystallinity is not impacted in *Region III* for a depth of more than 800 nm. For the first time, we elucidated that the $Gd_2Zr_2O_7$ waste form, which was considered to be structurally intact at 100 dpa, is completely amorphized at 6.5 dpa with the synergistic displacement damage, electronic energy loss, and He concentration enabled. This study leads to new physical insights into amorphization and He bubbles formation mechanisms of $Gd_2Zr_2O_7$ waste form under multi-energy He irradiation, which is essential for the design and optimization of irradiation-resistant ceramic waste matrices.

Gadolinium zirconate ($Gd_2Zr_2O_7$) has great potential to be applied as a waste form material to immobilize high-level radioactive wastes because of its high irradiation resistance and high actinide capacity. [1-5] After absorption of the energy from high-energy ions, $Gd_2Zr_2O_7$ rearranges the Gd, Zr, and O atoms within its crystalline structure enabling high stability under the radioactive repository conditions. [6] Multiple studies elucidated that under He irradiation the phase transition of $Gd_2Zr_2O_7$ from pyrochlore to defect fluorite was observed without experiencing complete amorphization. [7-9] Notably, the disordered defect fluorite $Gd_2Zr_2O_7$ presents high radiation resistance, which cannot be completely amorphized even at a high He dose of ~100 dpa. [1] Owing to such high irradiation resistance to amorphization, $Gd_2Zr_2O_7$ was proposed as a waste form for plutonium (Pu) disposition. [10] Moreover, $Gd_2Zr_2O_7$ can effectively immobilize a spectrum of radioactive nuclides, ranging from single component modeled compound, to complex multicomponent mixtures with multivalent nuclides, including the final waste after partitioning high-level liquid waste (HLLW) by using the trialkyl phosphine oxides (TRPO). [2,4,11,12,13] Existing literature is rich in studies on the radiation resistance performance of pure $Gd_2Zr_2O_7$ under He beam, however, few studies take the challenges for a systematic investigation into the structural and morphological evolutions of $Gd_2Zr_2O_7$ waste forms under He irradiation, because of the intrinsic complexity of the systems.

Mechanistic studies on the structural and morphological evolutions of nuclear waste forms under the radioactive environments of a geological repository are crucial for the design and optimization of stable waste forms for nuclear materials. Typically, ion implantation is used to simulate the irradiation effects, in which the interactions between ions and the solid lead to both nuclear and electronic energy losses. [14] The displacement damage from nuclear energy loss is caused by α-decay of the actinides, [15,16] which causes phase transition[3] and amorphization. [17] Meanwhile, the electronic energy loss also significantly influences the structures and morphologies of solid-state materials. For example, it has been reported by Zhang *et al.* that the electronic energy loss effectively annealed pre-existing defects and restored the structural order of silicon carbide. [18] Huang *et al.* elucidated that the thermally-activated ionization induced rapid amorphization of $Gd_2Zr_2O_7$ at low temperatures. [19] In addition, the local enrichment of He atoms from α-decay caused swelling and cracking of waste forms leading to potential mechanical degradation and increased leaching rate. [20,21] Thus far, studies on He irradiation of waste forms primarily focus on either displacement damage, electronic energy loss, and/or He accumulation. The understanding of their synergistic or integrated effects on structural and morphological mechanisms, which are tightly related to the stability of nuclear waste forms in the practical storage environment, is far from complete. Hence, systematic studies on the integrated impacts from both nuclear and electronic damages are needed to enhance our fundamental understanding of the prediction, design, performance assessment, and lifetime extension of $Gd_2Zr_2O_7$-based waste forms.

In this study, $Gd_2Zr_2O_7$ with TRPO waste (50 wt%) immobilized was irradiated with a He ion beam. The sample preparation processes, including material synthesis and irradiation, were described in detail in our previous studies. [5] Multi-energy He ion irradiation was applied to achieve near-uniform He concentration and displacement damage distribution, in which the degree of irradiation (doses) was determined by the SRIM-2008 full-cascade simulation code described in our earlier study. [22] Additionally, the electronic energy loss varies as the He ion penetration changes. Transmission electron microscopy (TEM) and fast Fourier transform (FFT) experiments were conducted on the post-irradiation $Gd_2Zr_2O_7$ samples to elucidate the phase transition, microstructural evolutions, and He bubble formation.

SRIM-2008 full-cascade simulation code was also used to determine the He irradiation parameters as described in our earlier study. [22] Based on the calculated ion concentration, a displacement per atom (dpa), and electronic energy loss, the $Gd_2Zr_2O_7$-based waste form prepared was irradiated under He ion beam at 10, 30, 70, and 140 keV to achieve a series of samples with different fluences, $4 \times 10^{16}$, $9 \times 10^{16}$, $1.2 \times 10^{17}$, and $2 \times 10^{17}$ ion/cm$^2$ at a dose rate of $1 \times 10^{13}$ ion/s·cm$^2$, respectively. The dose of He irradiation is considered during the simulation of electronic energy loss. [23] The He irradiation experiments were performed at room temperature using an ion implanter at the Ion Beam Application Center in the School of Physics at Wuhan University. The microstructure and morphology of each sample were studied using a Tecnai G2 F20 S-TWIN TEM operated at 200 kV. The samples for TEM experiments were prepared with a standard lift-out method using an FEI-Nova NanoLab 200 Focused Ion Beam (FIB).

The depth profiles of implanted He concentration, displacement damage, and electronic energy loss were calculated using SRIM simulation (see **Figure 1**). Within 500 nm below the surface exposed, multi-energy He irradiation leads to an average distribution of He concentration at about 10 at%, and an averaged damage dose of ~ 6.5 dpa. This suggests that multi-energy He irradiation can avoid the local stress at the interfaces of the irradiated and non-irradiated $Gd_2Zr_2O_7$ waste form, which is induced by the non-uniform implantation far from the surface irradiated. [24] In general, the electronic/ionization energy loss increases as the incident ion's energy increases and as the depth of irradiation decreases (see **Figure 1c**).

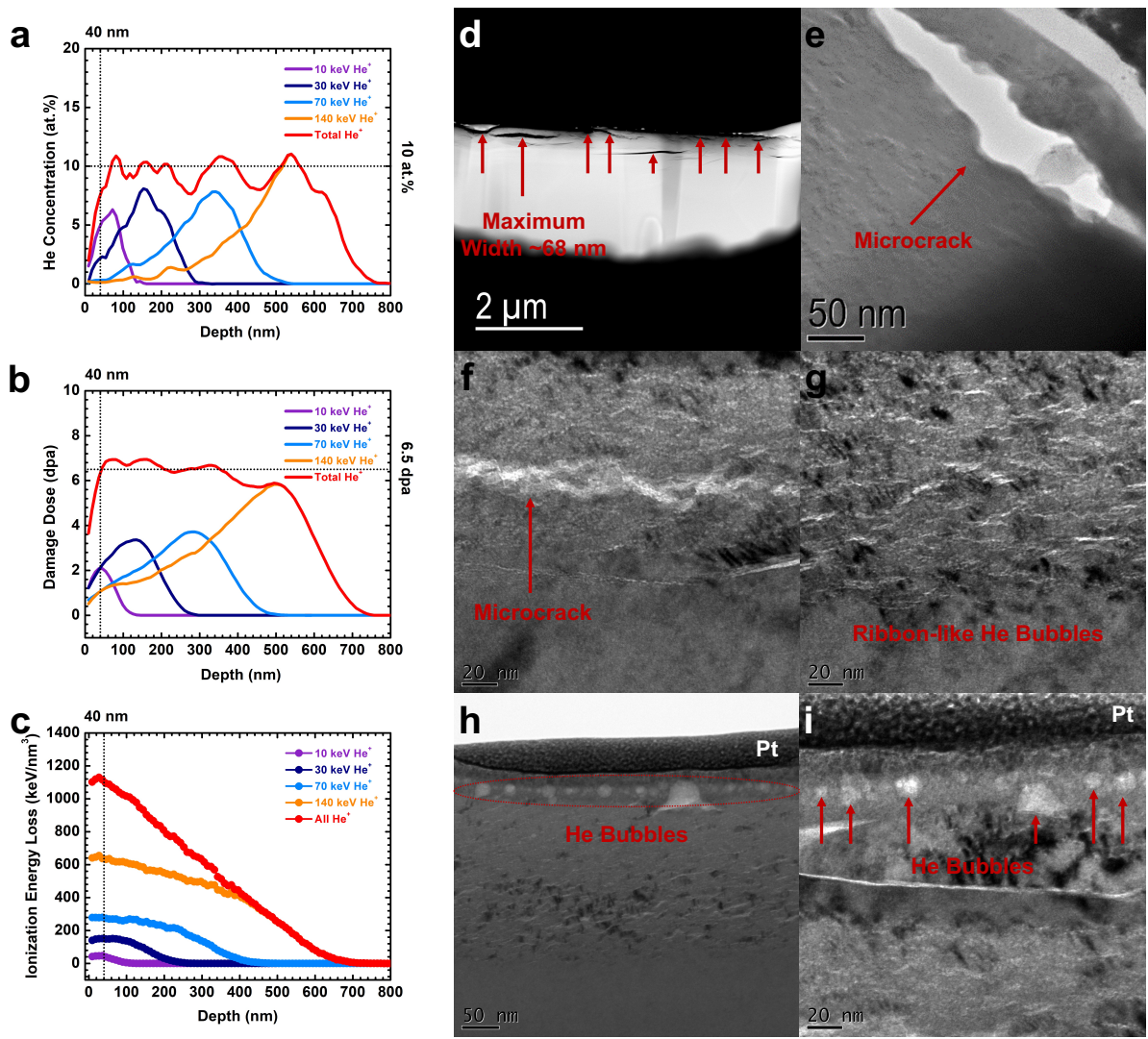

**Figure 1.** The simulation results using SRIM-2008 with the multi-energy He ion implantation. **(a)** He concentration, **(b)** displacement per atom (dpa), and **(c)** electronic energy loss as a function of depth (nm). The dashed line highlights the critical depth of complete amorphization (~ 40 nm). Complete amorphization may not occur at a distance further from the irradiated sample surface. TEM images of the $Gd_2Zr_2O_7$ waste form sample after multi-energy He ion irradiation: **(d)** the overview. **(e)**, **(f)** and **(g)** highlight different regions of **(d)** at the nanoscale. The TEM images of the near-surface regions of the $Gd_2Zr_2O_7$ waste form sample after multi-energy He ion irradiation are presented in **(h)** and **(i)**, the scale bars of which are 50 and 20 nm, respectively. The Pt clad layer is above the red dash line.

After multi-energy He irradiation, the microstructure of the sample exhibits a significant number of microcracks with width ranging from 10 to 68 nm (see the arrows in **Figure 1d**). The microcrack regions in **Figure 1d** are zoomed in and highlighted in **Figure 1e–g**, in which *four types* of microstructural defects were observed. Specifically, **Figure 1e** presents a completely cracked region of the He-irradiated $Gd_2Zr_2O_7$ waste form showing microcracks wider than 30 nm (*Type I*). Meanwhile, smaller microcracks sharing an average width of ~10 nm (*Type II*) were observed in **Figure 1f**. Interestingly, the *Type II* microcracks appear not to be completely developed from the ribbon-like He bubbles (*Type III*) shown in **Figure 1g**. Considering the dominant number of *Type II* microcracks, it is likely they are in the critical state towards the transition from ribbon-like He bubbles (*Type III*) to larger microcracks (*Type I*).

The TEM images of the near-surface regions of the $Gd_2Zr_2O_7$ waste form sample after multi-energy He ion irradiation are presented in **Figures 1h** and **i**. Interestingly, there are a significant number of well-aligned large spherical He bubbles (*Type IV*) at a depth of ~40 nm from the sample surface (see **Figure 1h**). In addition, large spherical He bubbles are only located within the depth less than 40 nm, while both ribbon-like He bubbles (*Type III*) and microcracks around 10 nm (*Type II*) appear more than 40 nm away from the He-irradiated surface. Interestingly, we notice that the near-surface He bubbles have diameters between 5 and 25 nm, which are much larger than those that reside inside the grains observed in multiple earlier studies one pure $Gd_2Zr_2O_7$.[3,25-27] Such a phenomenon strongly suggests that the formation of large spherical He bubbles in $Gd_2Zr_2O_7$ has been edited or altered by the immobilized TRPO waste (50 wt%).

HRTEM and fast Fourier transform (FFT) images of selected near-surface regions of the He irradiated $Gd_2Zr_2O_7$ waste form sample are presented in **Figures 2a** and **b**. At a depth less than 40 nm from the $Gd_2Zr_2O_7$ waste form surface (see **Figure 2a**), complete amorphization with the presence of large spherical He bubbles is identified. The absence of diffraction patterns in the FFT image further confirm that the *near-surface region* (*Region I*) with a depth less than 40 nm is completely amorphized by He ion irradiation. **Figure 2b** suggests that at a depth from 40 to 800 nm, the *intermediate region* (*Region II*) with a significant number of large spherical He bubbles, is overlapped with the completely amorphized *near-surface region* (*Region I*). The diffraction patterns in the FFT image suggest this *intermediate region* is partially amorphous. More specifically, the sample is more disordered close to *Region I* (see the *orange square* in **Figure 2b**) and has higher crystallinity at a further distance up to 800 nm (see the *green square* in **Figure 2b**). Pyrochlore $Gd_2Zr_2O_7$ structure with high crystallinity (crystalline region, *Region III*) is observed in the region far from the sample surface, with a depth more than 800 nm, directly exposed to He irradiation (see **Figures 2c**). Therefore, based on the experimental evidence, we successfully identify that the critical amorphization depth is around 40 nm from the sample surface exposed to He irradiation, where the He atom concentration, damage dose, and electronic energy loss are 8 at%, 6.5 dpa, and 1100 $keV/nm^3$, respectively (see the dash lines in **Figure 1**).

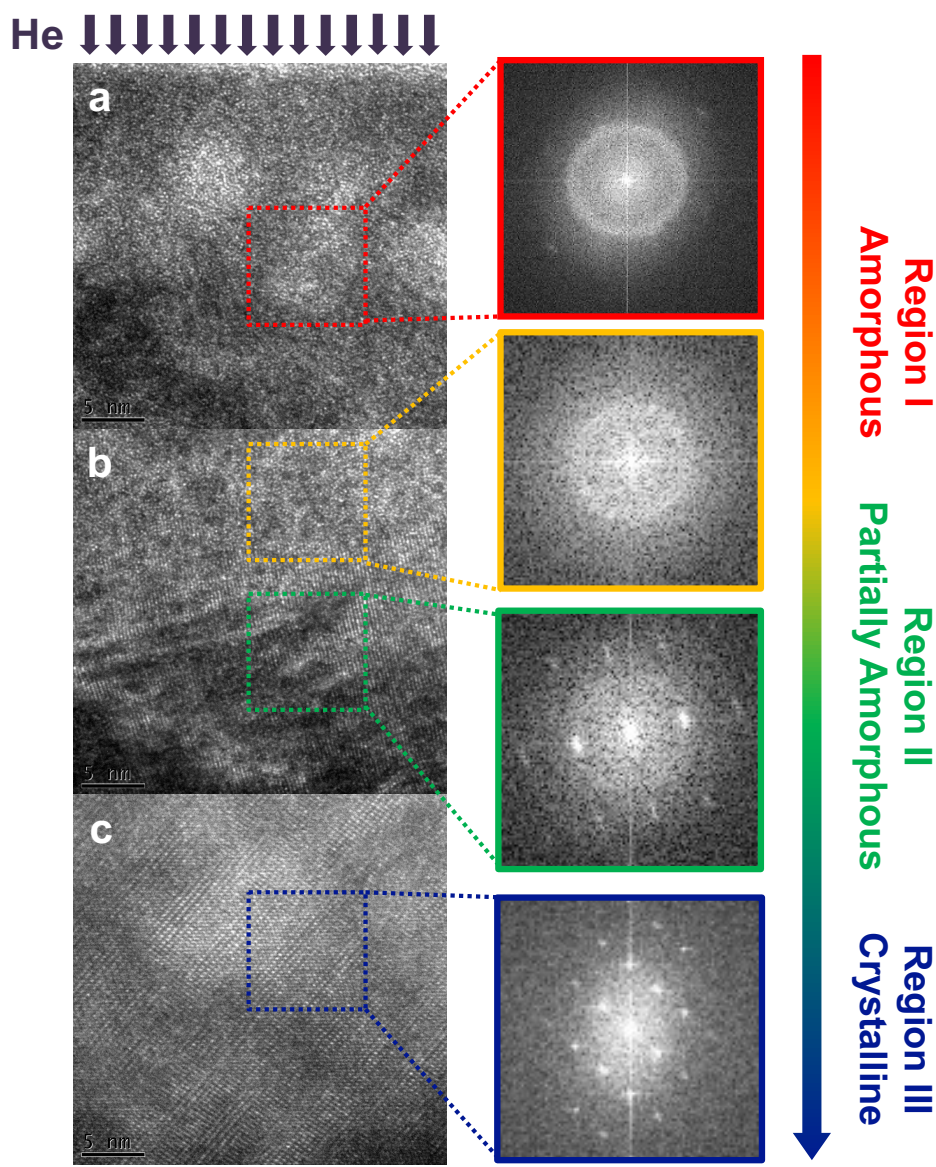

**Figure 2.** HRTEM images of He-irradiated $Gd_2Zr_2O_7$ waste form with TRPO (50 wt%), and the fast Fourier transform (FFT) images of highlighted regions: **(a)** *Region I* – within 40 nm from the He-irradiated surface (red frame), **(b)** *Region II* – between 40 and 800 nm. The orange frame highlights the transition from the highly amorphous *Region I* to a mildly amorphous region closed to 40 nm, and the green frame shows the region with higher crystallinity. **(c)** *Region III* – the crystalline region more than 800 nm away from the He-irradiated surface (blue frame).

According to the TEM results, we conclude that the integration of high electronic energy loss, high He atom concentration, and displacement damage leads to synergetic effects that substantially accelerate the amorphization of the polycrystalline $Gd_2Zr_2O_7$ ceramic sample. Zhang *et al.* reported that He irradiation of $C_{60}$ clusters at 12-MeV and $1\times10^{11}$ $ions/cm^2$ led to an electronic energy loss of 32 keV/nm, and the sub-surface layer of $Gd_2Zr_2O_7$ pyrochlore was significantly damaged and exhibited coexisted nanocrystalline defect-fluorite and amorphous phases. [28] The formation of such a liquid-like amorphous phase under irradiation is typically attributed to significant electronic energy loss. [14,29] However, in our earlier report, although the ionization energy applied on $Gd_2Zr_2O_7$ exposed to He irradiation (190 keV and $5\times10^{17}$ $/cm^2$) was higher than what we used in the current study, no complete amorphization was observed. [25] Therefore, it is likely that high electronic energy loss may not be the single dominant factor that governs complete amorphization, and the He atom concentration and displacement damage may also play critical roles. More specifically, for $Gd_2Zr_2O_7$ although He irradiation-induced displacement damage interrupted the crystallinity, the simultaneous rearrangement of lattice atoms structurally inhibits such effect.[6] However, the high concentration of insoluble He (bubbles) may inhibit the atomic rearrangement, resulting in reduced resistance to irradiation-induced amorphization. We would like to emphasize that based on our observation neither high He concentration nor displacement damage leads to complete amorphization. For example, even at a high damage dose of 100 dpa, $Gd_2Zr_2O_7$ preserves a partially ordered crystal structure.[1] Moreover, $Gd_2Zr_2O_7$ retains a significant portion of crystallinity without a complete amorphization at a high He atom concentration of ~20%.[25]

Interestingly, we observed large spherical He bubbles with diameters ranging from 5 to 25 nm in the completely amorphous region. Such phenomenon was not seen on pure $Gd_2Zr_2O_7$ under He irradiation. The point defects generated by irradiation, such as vacancies and interstitials, feature different mobility behavior. Specifically, the fast-moving interstitial He atoms preferentially annihilate at the free surface, resulting in an increase in slower-moving vacancies near the surface.[30] High electronic ionization loss induces the migration and coalesces of vacancies towards the formation of larger voids.[31] Such large voids are energetically metastable until the capture and encapsulation of He atoms,[32] leads to the formation of a completely amorphous phase with homogeneous isotropic medium and spherical He bubbles, which feature minimized surface energy to reach an energetically favorable state.[33] This means that in the interplays between surface energy and elastic strain energy, the former plays a dominant role in the determination of the *completely amorphous region* (*Region I*),[26] while the grain boundary is anisotropic. As the He bubbles nucleate and grow at the grain boundaries, the elastic strain energy of He bubbles growing perpendicular to the grain boundary is much higher than that of He bubbles growing along the grain boundary. Therefore, the shapes of He bubbles governed by elastic strain energy appear to be ribbon-liked residing along the grain boundary. [34] In addition, based on the elastic theory, [35] the maximum radius of the spherical He bubbles ($R_m$) can be calculated using **Equation 1**.

$$R_m = 8\mu\gamma/P^2_{eff} \qquad \textbf{\textit{Equation 1}}$$

where $\mu$ is the shear modulus of $Gd_2Zr_2O_7$, $\gamma$ is the surface tension constant, and the effective pressure of He bubbles is $P_{eff}$. [33]

$$P_{eff} = P_{surface} + P_{He} \qquad \textbf{\textit{Equation 2}}$$

$P_{surface}$ and $P_{He}$ are the interfacial tension, and the internal He pressure, respectively, where $P_{He}$ is obtained from the He density inside the bubble. Typically, $P_{surface}$ is omitted, compared with the quite high $P_{He}$ that may keep He in solid-state at room temperature. [26,33] Thus, $P_{eff}$ is proportional to $P_{He}$, the density of He in bubbles that is positively correlated with the He concentration. According to **Equation 1**, as the He concentration, $R_m$ decreases. In the region completely amorphized (*Region I*), the He concentration calculated by SRIM is below 8.0 at%, which is lower than that within the grain. Thus, the $R_m$ magnitude of the spherical He bubbles is much higher, which is consistent with the observation that the larger spherical He bubbles are typically detected in *Region I*.

For the region slightly deeper than 40 nm from the surface (*Region II*), partial amorphization is observed associated with ribbon-like He bubble (*Type III*) and small microcracks narrower than 10 nm (*Type II*). At a depth significantly further than 40 nm away from the surface irradiated, the electronic energy loss simulated by SRIM significantly and sharply decreases. Hence, complete amorphization cannot be achieved. The partially amorphized region (*Region II*) is anisotropic, resulting in the formation of ribbon-like He bubbles (*Type III*). Furthermore, as the He concentration increases, a large number of atoms belonging to the $Gd_2Zr_2O_7$ waste form are ejected and become interstitial atoms, leading to a rapid increase in the elastic strain energy. [26] The ribbon-like He bubbles tend to appear at higher He concentrations because their shapes are determined by the elastic strain energy. Further increase in the concentration of He may lead to growth and coalescence of the ribbon-like He bubbles and may eventually result in the formation of microcracks to minimize the elastic strain energy. [27]

In this study, we investigated the microstructural and helium bubbling evolutions of $Gd_2Zr_2O_7$ waste form with 50 wt% of TRPO immobilized under multi-energy He ion irradiation. Within 40 nm below the He-irradiated surface (*Region I*), the $Gd_2Zr_2O_7$ waste form is completely amorphized. Such amorphization is induced by the synergetic effects of integrated high electronic energy loss, high He atom concentration, and displacement damage. Moreover, the completely amorphized phase in *Region I* is accompanied by large spherical He bubbles (*Type IV*) with diameters ranging from 5 nm to 25 nm. At a depth between 40 and 800 nm from the He-irradiated surface (*Region II*), the $Gd_2Zr_2O_7$ waste form is partially amorphous with ribbon-like He bubbles (*Type III*) enabled by increased He local concentration. Further growth and combination of such ribbon-like He bubbles lead to the formation of microcracks. The sample studied is fully crystalline at a depth of more than 800 nm (*Region III*). The new phenomena observed here have enhanced our understanding of (i) amorphization and (ii) He bubbles formation mechanisms of $Gd_2Zr_2O_7$ waste form under multi-energy He irradiation.

**Data Availability Statement**

The data that support the findings of this study are available from the corresponding authors upon reasonable request.

**Acknowledgments**

This work was supported by the National Natural Science Foundation of the People's Republic of China under Grant No. 11775152, the Key Research and Development Program of Sichuan Provence under Grant No. 2020YFG0192. Di Wu thanks the institutional funds from the Gene and Linda Voiland School of Chemical Engineering and Bioengineering and Alexandra Navrotsky Institute for Experimental Thermodynamics at Washington State University.